\documentclass[aps,prl,showpacs,twocolumn,amsmath,amssymb,footinbib]{revtex4}

\usepackage[english]{babel}
\usepackage{latexsym}
\usepackage{graphics}
\usepackage{subfigure}
\usepackage{epsfig}
\usepackage{amsfonts}
\usepackage{amssymb}
\usepackage{amsmath}
\usepackage{bbm}

\begin{document}

\title{Mott-insulator states of ultracold atoms
in optical resonators}

\author{Jonas Larson$^1$, Bogdan Damski$^2$, Giovanna Morigi$^3$, and Maciej Lewenstein$^{1,4}$}
\address{$^1$ ICFO--Institut de Ci\`encies
Fot\`oniques, E-08860 Castelldefels, Barcelona, Spain\\
$^2$ Theory Division, Los Alamos National Laboratory, MS-B213, Los Alamos, NM 87545, USA\\
$^3$ Departament F\' isica, Grup d'\'Optica, Universitat Aut\'onoma de Barcelona, E-08193
Bellaterra, Spain\\
$^4$ ICREA--Instituci\'o Catalana di Recerca i Estudis Avan{\c c}ats, Pg Llu{\'i}s Companys 23, E-08010 Barcelona, Spain}  \date{\today}

\begin{abstract} We study the low temperature physics of an ultracold
atomic gas in the potential formed inside a pumped optical
resonator. Here, the height of the cavity potential, and hence the
quantum state of the gas, depends not only on the pump parameters,
but also on the atomic density through a dynamical a.c.-Stark
shift of the cavity resonance. We derive
the Bose-Hubbard model in one dimension,  and use the strong coupling expansion to determine the parameter regime in
which the system is in the Mott-insulator state. We predict the
existence of overlapping, competing Mott states, and bistable
behavior in the vicinity of the shifted cavity resonance,
controlled by the pump parameters. Outside these parameter
regions, the state of the system is in most cases superfluid.
\end{abstract}

\maketitle

Ultracold atomic gases in optical lattices offer the unprecedented
and unique possibility to study paradigmatic systems of quantum
many-body physics~\cite{bloch,ml}. These systems allow one to
realize various versions of Hubbard models \cite{zoller}, a
prominent example of which is the Bose-Hubbard model
\cite{fisher}, exhibiting the superfluid (SF) -- Mott insulator
(MI) quantum phase transition \cite{sachdev}. The realization of
the Bose-Hubbard model with ultracold atoms has been proposed in
the seminal theoretical work in Ref.~\cite{jaksch},
and demonstrated in the milestone experiments in
Ref.~\cite{blochex}. Several aspects and modifications of the SF -- MI
quantum phase transition (or crossover \cite{batroumi}) are object
of intense studies~\cite{ml}.

Optical lattices in free space are not affected by the presence of the atoms. This scenario is, however, strongly modified when the atoms move in the optical potential which is formed inside a pumped resonator: Here, the atoms interact with the
cavity mode while the cavity field, determining the optical
lattice, may critically depend on the density of the
atoms \cite{JOSA_Helmut,Footnote:1}. Several recent studies address 
Cavity Quantum Electrodynamics (CQED) with cold atoms. CQED techniques were used to measure pair
correlations in the atom laser \cite{esslinger}, and have been proposed for characterizing quantum states of ultracold matter~\cite{HelmutNature}. Self-organization
of atoms in transversally pumped cavities was
observed in~\cite{black}, and theoretically described
in~\cite{asboth}. Bragg scattering of atomic structures inside
optical resonators has been investigated in~\cite{Zimmermann}.
Most recently, Bose-Einstein condensed atoms have been loaded
inside cavities~\cite{BEC-load}. This experimental progress calls
for theoretical development of CQED combined with many-body
physics.

In this Letter we determine the ground state of ultracold atomic
gases in the optical lattice of a cavity. The cavity is driven by
a laser, and the atoms shift the cavity resonance, thus affecting
the intracavity field amplitude, which in turn determines the
depth of the cavity potential and the ground state of the atomic
gas itself. The problem is hence highly non-linear, as the optical
lattice and state of the atoms have to be evaluated in a
self-consistent way. The derivation of the corresponding
Bose-Hubbard model for few atoms has been discussed by Maschler
and Ritsch in Refs.~\cite{Maschler}. In this Letter, we derive the Bose-Hubbard model in an appropriate 
thermodynamic limit. We study its ground state applying the strong
coupling expansion~\cite{monien} to calculate the boundaries of
the MI states, determined by the dependence on the parameters of
the system: pump strength and frequency, density of atoms, and
chemical potential.

Our model consists of bosonic atoms confined in a 1D trap inside
an optical resonator of a fixed length driven by a laser field. The atomic
dipole transition is far-off resonance from the cavity mode, which
induces a dipole potential acting on the atoms. Using the notation
of~\cite{Maschler}, the single-particle Hamiltonian reads:
\begin{equation}\label{H:0} \hat{H}_0=\frac{\hat{p}^2}{2m}+\hbar
\left[U_0\cos^2(k\hat{x})-\Delta_c\right]\hat{n}_{\rm
ph}-i\hbar\eta\left(\hat{a}-\hat{a}^\dagger\right). \end{equation}
Here, $\hat{p}$, $\hat{x}$ and $m$ are the atomic momentum,
position, and mass, $\eta$ is the amplitude of the pump at
frequency $\omega_p$, $\Delta_a=\omega_p-\omega_a$ and
$\Delta_c=\omega_p-\omega_c$ are the detunings of the pump from
atom and cavity frequencies, $k=\omega_c/c$ is the mode wave
vector, $\hat{a}^\dagger$ and $\hat{a}$ are the creation and
annihilation operators of a cavity photon of energy
$\hbar\omega_c$, and $\hat{n}_{\rm ph}=\hat{a}^{\dagger}\hat{a}$
is the number of photons. The depth of the single-photon dipole
potential is $U_0=g_0^2/\Delta_a$, where $g_0$ is the atom-cavity
mode coupling. The many-body Hamiltonian is obtained from
Eq.~(\ref{H:0}) including the atomic contact interactions; it is conveniently represented  in second-quantized 
form with the atomic field
operators $\hat{\Psi}(x)$, $\hat{\Psi}^{\dag}(x)$ obeying the bosonic commutation
relations. We assume the
bad-cavity limit, where the resonator field reaches
the stationary state on a faster time scale than the one of the
atomic dynamics, and eliminate the cavity field from
the equations of the atomic operators. In this limit the amplitude
of the intracavity field depends non-linearly on the atomic
fields through the operator $\hat{\mathcal Y}=\int{\rm
d}x\cos^2(kx)\hat\Psi^{\dagger}(x)\hat\Psi(x)$, and reads
\begin{equation} \hat a(\hat{\mathcal Y})=\frac{\eta}{\kappa-{\rm
i}(\Delta_c-U_0\hat{\mathcal Y})},
\label{F} \end{equation} where $\kappa$ is the cavity damping
rate. Correspondingly, the Heisenberg equation for the atomic
field operator reads \begin{eqnarray} \label{Eq:Psi}
\dot{\hat{\Psi}}&=&-\frac{\rm i}{\hbar}[\hat\Psi(x),\hat{\cal
H}_0]-{\rm i}\hat{\mathcal C}(\hat{\mathcal Y},x), \end{eqnarray}
where $\hat{\cal H}_0=\int\!dx\,\hat\Psi^{\dag}(x)
\left(-\frac{\hbar^2\nabla^{2}}{2m}+\frac{u}{2}\hat\Psi^{\dag}(x)\hat\Psi(x)\right)\hat\Psi(x)$,
with $u$ being the strength of the contact interaction, and
\begin{eqnarray} \hat{\mathcal C}(\hat{\mathcal
Y},x)=U_0\cos^2(kx)\hat a^{\dagger}(\hat{\mathcal
Y})\hat\Psi(x)\hat a(\hat{\mathcal Y})\label{C:eff}\end{eqnarray}
which arises from keeping track of the correct normal ordering of
atomic and photonic field operators. Starting from Eq.~(\ref{Eq:Psi}), the derivation of the
Bose-Hubbard Hamiltonian is not straightforward due to the form of operator~(\ref{C:eff}).
This is evident when expanding the atomic field
operators, assuming the validity of the tight-binding
approximation (TBA) and the occupation of the lowest energy band:
$\hat{\Psi}(x)=\sum_iw(x-x_i)\hat{b}_i$, where $\hat{b}_i$ and
$w(x-x_i)$ are the atomic annihilation operator and Wannier
function at site $i$, respectively. The Wannier functions depend on $\hat{\mathcal
Y}$ and, therefore, on the number of atoms $N$ or, equivalently, on the atomic density, 
which in turn depends on Wannier functions.
Moreover, the
commutation relation $[\hat{b}_i,\hat{b}_j^{\dagger}]=\delta_{ij}$
is valid only in the  lowest order in the expansion in $1/N$. In effect, the Wannier
expansion must be  performed self-consistently in the thermodynamic limit, by letting $N$ and the cavity volume to infinity, keeping finite the number of atoms  per potential site. Additionally, we impose 
$U_0=u_0/N$ and $\eta=\eta_0 \sqrt{N}$, where $u_0$ and $\eta_0$ are constants, which corresponds to keeping
the depth of the cavity potential $V=\hbar U_0 n_{\rm
ph}$ constant as $N$ is increased. The Bose-Hubbard Hamiltonian $\hat{H}$ is obtained 
discarding couplings beyond nearest-neighbor. Its
rescaled form $\hat{\tilde{H}}=\hat H/U$, with $U$ the strength of the
on--site interaction, reads \begin{equation}\label{effham3}
\hat{\tilde{H}} =
-\tilde{t}\hat{B}+\frac{1}{2}\sum_i\hat{n}_i(\hat{n}_i-1)-\tilde{\mu}
\hat{N}, \end{equation} where
$\hat{N}=\sum_i\hat{n}_i=\sum_i\hat{b}_i^\dagger \hat{b}_i$ is the
atom number operator and $\hat{B}=\sum_i \hat{b}_i^\dagger
\hat{b}_{i-1}+\mathrm{h.c.}$ is the hopping term. The term $\tilde{\mu}=\frac{\mu}{U}+\frac{f(\hat{N})}{\hat{N}U}$
contains the rescaled chemical potential, where the second term is
a constant in the thermodynamic limit. The tunneling
parameter \begin{eqnarray}\label{eq:t}
\tilde{t}&=&-\frac{E_{1}}{U}-\frac{\hbar\eta^2
U_0J_{1}}{U\left(\kappa^2+\zeta^2\right)} \end{eqnarray} is
expressed in terms of
$\zeta=\Delta_c-u_0J_0\hat{n}_0$ and of the coefficients $U=u/2\int
dx\,|w(x)|^4$, $E_{\ell}=\int
dx\,w(x-x_l)(-{\hbar^2}/{2m})({d^2}/{dx^2})w(x-x_{l+\ell})$ and
$J_{\ell}=\int dx\,w(x-x_l)\cos^2(kx)w(x-x_{l+\ell})$, with
$\ell=0,1$. Note that $\hat N$, and hence the atomic density in the homogeneous case, is a conserved quantity since
$[\hat{N},\hat{H}]=0$. In deriving Eq.~(\ref{effham3}) we have used that
$J_1\ll J_0$. The higher order terms in $J_1\hat{B}$, describing
long-range interactions, have hence been neglected.  We note that the parameters $\tilde{\mu}$ and $\tilde{t}$ depend
on the atomic density through the Wannier functions, and at the same time determine the state of the system, and in particular the density: 
This is a genuine CQED effect, where the non-linearity of the coupling between photons and atoms depends on the atom number. As a consequence, the atomic density in this system is not determined by the chemical potential alone. 

From Eq.~(\ref{effham3}) we determine the parameter regimes of the
MI states using the strong coupling expansion~\cite{monien}. Here,
the boundaries of the MI regions are determined by comparing the
energy of the MI state, given by $n_0$ atoms at each site of the
periodic potential, with the corresponding energies of the excited
states with one additional or missing particle (particle and hole
states). This procedure involves the evaluation of the Wannier
functions for all cases. As the coefficients of the Bose-Hubbard
Hamiltonian both depend on and determine the atomic density, the
Wannier functions have to be calculated self-consistently. This is
done by solving the non-linear equation in presence of the
potential $V=\hbar U_0n_{\rm ph}=\hbar
u_0\eta_0^2/\left[\kappa^2+\left(\Delta_c-u_0J_0n_0\right)^2\right]$,
where $J_0$ is an integral of Wannier functions. This equation
cannot be solved by iteration, as one encounters periodic doubling
bifurcations and deterministic chaos. We solve it numerically by
checking for self-consistent solutions, using the Gaussian
approximation of the Wannier functions, and thus approximating
$w(x)\approx\exp(-x^2/2\sigma^2)/(\sqrt{\pi}\sigma)^{1/2}$ where
$\sigma$ is the parameter to be determined~\cite{footnote}. In terms
of the dimensionsless quantity $y=k^2\sigma^2$, giving the
extension of the Gaussian wave packet in units of the cavity mode
wavelength, the problem can be reduced to solving
self-consistently the equation $J_{0}(y)=\frac{1}{2}\left(1-d_a
e^{-y}\right)$, where $d_a$ is the sign of the detuning
$\Delta_a$. For a given set of parameters, multiple (bistable~\cite{Bonifacio})
solutions appear when the number of photons is maximum, namely
when the denominator of Eq.~(\ref{F}) is minimum, which occurs at
the shifted resonance \begin{equation}\label{zeros}
\Delta_c-u_0J_{0}n_0=0.\end{equation} Since the sign of $u_0$ is
determined by the detuning $\Delta_a$, Eq.~(\ref{zeros}) allows
for real solutions only when $\Delta_a$ and $\Delta_c$ have the same
sign: Then, the resonance condition depends on the number of atoms.
Correspondingly, the cavity is driven
at resonance, the number of photons reaches the maximum value
$n_{\rm ph}=\eta^2/\kappa^2$, and the cavity potential is the
deepest. An important distinction must be made between the cases
$\Delta_a>0$ ($U_0>0$) and $\Delta_a<0$ ($U_0<0$): In the first
case, the potential minima are at the nodes of the standing wave,
where the intracavity field vanishes. Strong localization of the
atoms at these points implies that the coupling of the atoms with
the field is minimum, $J_0\to 0$. The quantum fluctuations give
rise to a finite coupling, determining the quantum state. On the
contrary, when $\Delta_a<0$ the potential minima are at the
antinodes of the standing wave, where the intracavity field is
maximum. Strong localization of the atoms at these points implies
strong coupling with the field, with $J_0\to 1$. In this regime,
CQED effects are expected to play a dominant role.

We plot the boundaries of the resulting Mott states in the
$\tilde{\mu}-\eta^{-1}$ plane, i.e., the effective chemical
potential and the inverse of the pump strength. Here, large pump
strengths correspond to deep optical potentials, hence to
vanishing tunneling, $t\to 0$. The physical system we consider is
a gas of $^{87}$Rb atoms with scattering length $a_s=5.77$ nm,
whose dipole transition at wavelength $\lambda=830$ nm couples to
the mode of a resonator at decay rate $\kappa=2\pi\times100$ kHz.
The potential has transverse size $\Delta_y=\Delta_z=30$ nm and
$K$ sites in the longitudinal axis. We evaluate the "phase
diagrams" for $K=50-10000$ at fixed number of atoms $N$, scaling
$N$ so to keep the atomic density constant. The results for
the Mott zones agree over the whole range of values, so in the figures
we report the ones obtained for $K=50$.

\begin{figure}[th]
\includegraphics[width=0.6\linewidth]{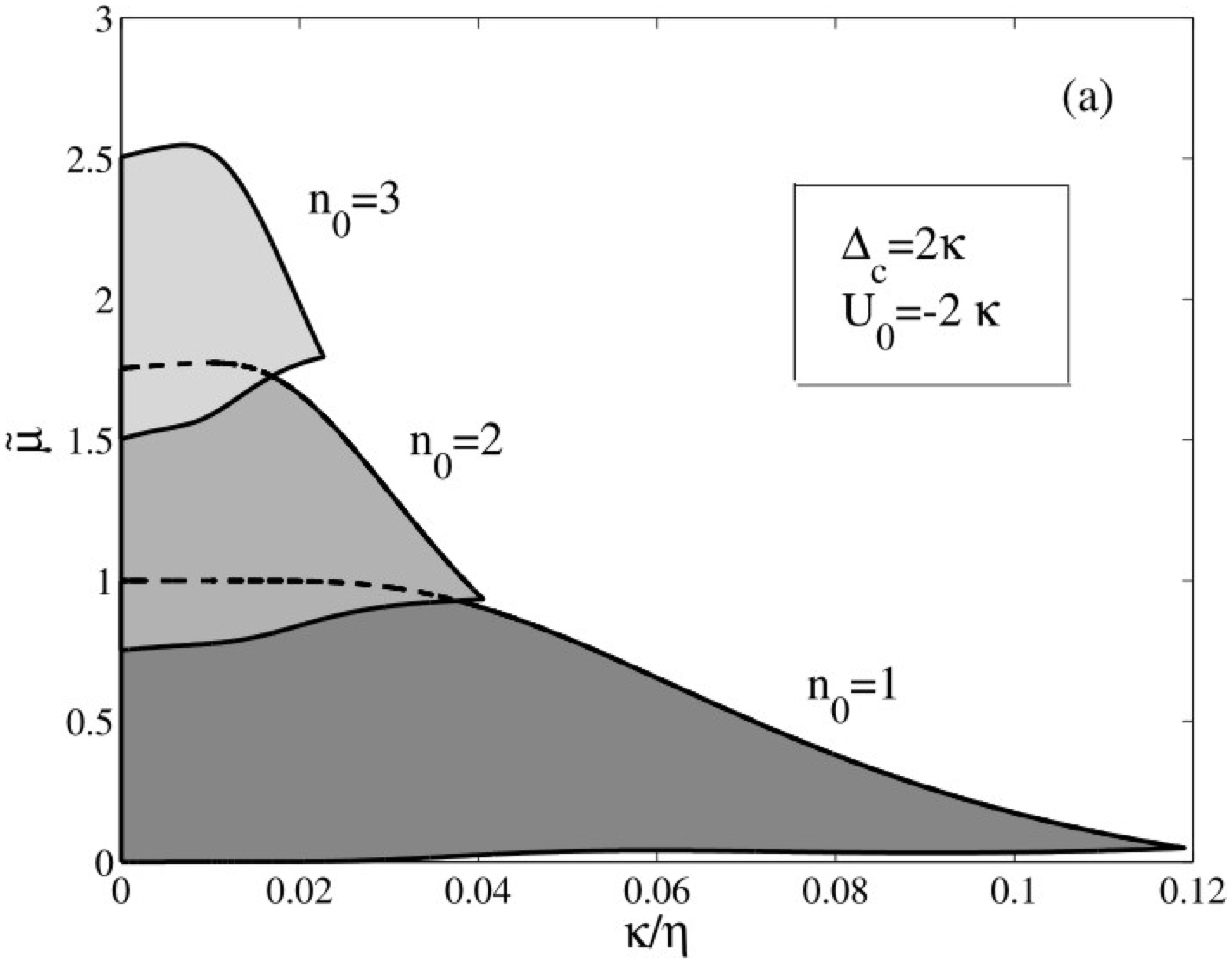}
\includegraphics[width=0.6\linewidth]{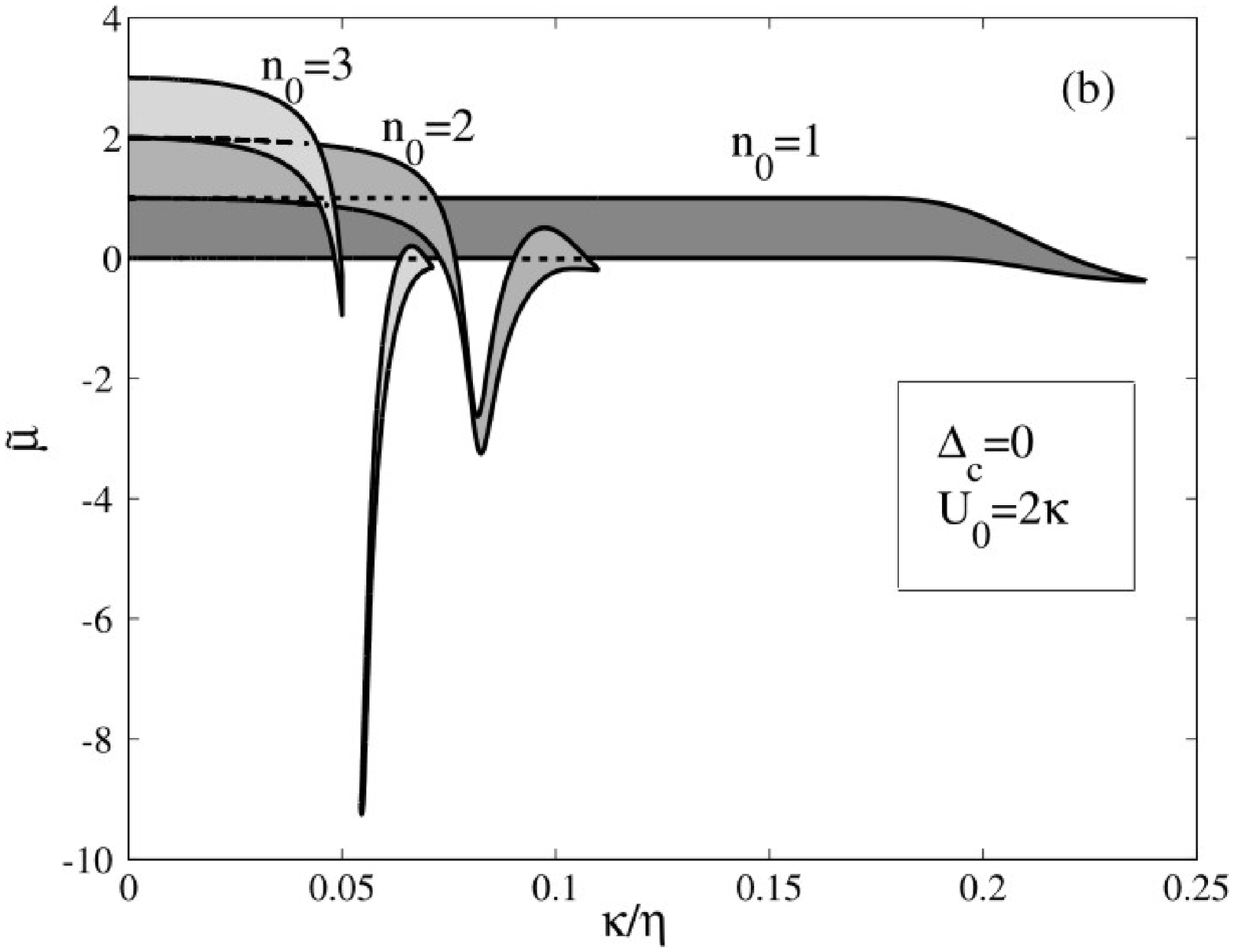} \caption{Boundaries of MI regions as a function of the 
rescaled chemical potential $\tilde \mu$ and the inverse of the pump strength $\eta$ (in units of $\kappa$) for (a)
$\Delta_c=2\kappa$ and $U_0=-2\kappa$ and (b) $\Delta_c=0$ and
$U_0=2\kappa$; $n_0$ denotes the site occupation in the 1D
cavity lattice potential of $K=50$ sites. The dashed lines show the boundaries of the zones which are hidden.} \label{fig1} \end{figure}

We first discuss the case in which the detunings $\Delta_a$ and
$\Delta_c$ have different signs, i.e., far from bistability~\cite{Bonifacio} when Eq.~(\ref{zeros}) is
not fulfilled. As expected, there is a peculiar difference between
the cases $\Delta_a>0$ (atoms at the nodes) and $\Delta_a<0$
(atoms at the antinodes): When $\Delta_a<0$, in the tight-binding regime $J_0\to  1$ and thus
$V\propto 1/(\kappa^2+(\Delta_c+n_0u_0)^2)$. Hence, 
the dependence on the atomic density is strongest. In
Fig.~\ref{fig1}(a) the MI zones are displayed. The shapes are
similar to the standard ones~\cite{monien}, apart from the region
$\eta\to\infty$ where, despite being out of the bistable regime~\cite{Bonifacio}, the lobes considerably overlap. This effect is due to the competition between the non-linear coupling to the cavity field, giving the depth of the potential, and the strength of the onsite interactions, affecting the number of atoms per site $n_0$. 
In the other case,
$\Delta_a>0$, one has $J_0\ll  1$. For $\eta\to\infty$, then $J_0\to 0$, the cavity potential depth
is almost independent of $n_0$ and one obtains the standard
Bose-Hubbard Hamiltonian. For large but finite values of $\eta$,
however, $J_0$ is finite and the dependence of the coefficients on $n_0$ becomes relevant. Figure~\ref{fig1}(b) shows the
``phase diagram'' for $|U_0|=2\kappa$ and $\Delta_c=0$. Here, the
MI regions exhibit a regular behavior at $\eta\to\infty$. As
$\eta$ is decreased they start to overlap and become disconnected. This behaviour introduces two new critical points at the tips of the disconnected regions, whose nature will be studied in future works.  We note that the MI zones enter the region of negative $\tilde{\mu}$. The minima of $\tilde{\mu}$
are at the pump values where the on-site repulsion is
balanced by the effective potential $V$.

\begin{figure}[th]
\includegraphics[width=0.6\linewidth]{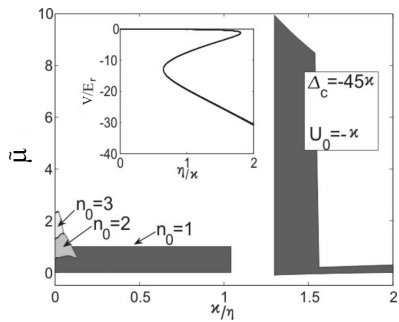} \caption{Phase
diagram showing the MI regions for
$\Delta_c=-45\kappa$ and $U_0=-\kappa$. Inset: $V$ (in units of
$E_r$) as a function of $\eta$ for $n_0=1$. The bistability causes the system to go out of
the left MI region with $n_0=1$. The right MI-region for $n_0=1$ is an artifact of
the theory~\cite{monien}, as here the TBA is invalid.
} \label{fig4} \end{figure}

We now consider the situation when $\Delta_a$ and $\Delta_c$ have
the same sign, such that Eq.~(\ref{zeros}) may have real
solutions. From Eq.~(\ref{zeros}), for $n_0=1$ and $U_0=-\kappa$ we find bistable behavior
for $J_0=|\Delta_c|/50\kappa$ and $J_0$ sufficiently close to 1,
which is fulfilled for instance for $\Delta_c=-45\kappa$. The
corresponding diagram is displayed in Fig. \ref{fig4}. The inset
shows the potential $V$ as a function of $\eta$ for $n_0=1$. Here, one encounters a bistability
point while lowering the pump intensity, where the cavity field potential discontinuously jumps to
a second branch with $|V|\ll E_r$, $E_r=\hbar^2k^2/2m$ being the
recoil energy. The first branch corresponds to the left
MI-region of the phase diagram at $n_0=1$. In the second branch, instead, the TBA is not valid, hence
most probably the atomic gas will no longer be in the lowest band
of the cavity potential, and rather definitely  no longer in a MI
state. Using both Gaussian and Wannier functions we have verified
that our treatment breaks down as soon as the system goes out from
the MI region on the left of Fig.~\ref{fig4}. This instability has
the character of a first order transition. The right MI region in
Fig.~\ref{fig4} is found by applying the theory of~\cite{monien}.
It occurs at values of $\eta$ for which $|V|$ is in the second
branch, and is thus of dubious validity, since here the TBA breaks
down. Instability leads here apparently to 
population of higher Bloch bands; most probably the true
ground state in this regime is SF (BEC in a very weak lattice
potential). Let us conjecture that we could find parameter regions where the
cavity potential for both branches of solutions would support the TBA.
In that case, at a given density there would exist two stable values of
the tunneling and onsite-interaction matrix elements. Then, for a
fixed $\tilde{\mu}$ we would have two possible phases, of which only one
will be energetically favorable, but both being by construction stable with respect 
to small perturbations, such as single particle or hole excitations.

So far we have considered a fixed number of atoms. If instead the system 
is coupled to an atomic reservoir, and the number of
atoms is hence not fixed, then a change of the system parameters
can lead either to continuous or to abrupt changes of the atomic
density, and hence of the ground state, analogue to second- and
first-order phase transitions, respectively. When the parameters
are such, that the system does not exhibit bistability, then $\tilde{\mu}$, $\kappa/\eta$ and the atomic
density determine uniquely the coefficients in the Bose-Hubbard
model and the ground state. In the case of overlapping MI zones, 
as in Fig.~\ref{fig1}(a), the system would relax to the state with the density which minimizes the energy. 
Nonetheless,
by slowly changing $\tilde{\mu}$, corresponding to spanning the phase diagram along the ordinate,
we expect hysteresis in the atomic on-site density, since the states inside the Mott correspond to local energy minima.  
Note that such local changes of on-site $\tilde{\mu}$ are also encountered when the system is
inhomogeneous, for instance, in presence of a harmonic trap. The atomic density will exhibit then the characteristic 
"wedding cake" form (cf. \cite{ml,jaksch,batroumi}). Sufficiently slow changes of the trap frequency should then lead to hysteresis in the "wedding-cake" shape. Applications of this effect could include many-body quantum switches and generation of coherent superpositions of Mott states. 

This letter refers to the case, in which the atomic density globally affects the cavity field. Situations,
when the atoms may affect locally the cavity field, can be found in multimode resonators~\cite{marciek,Meystre}. In these
scenarios one could find features typical of phonon-like physics in solid state. We acknoledge discussions with I. Bloch, T. Esslinger, S. Fernandez, Ch. Maschler, H. Monien, E. Polzik, J. Reichel, and H. Ritsch, and support from the Swedish
government/Vetenskapsr{\aa}det, the German DFG (SFB 407, SPP
1116), the EU commission (SCALA, Contract No.\ 015714), ESF PESC
QUDEDIS, the US Department of Energy, and the Spanish MEC
(FIS2005-04627, Ramon-y-Cajal, Consolider Ingenio 2010 "QOIT").


\begin{thebibliography}{999}
\bibitem{footnote} We use modified functions
from a Gaussian ansatz, satisfying orthogonality
between neighbouring sites.
\end{thebibliography}
\end{document}